\documentclass[12pt]{iopart}
\usepackage{iopams}  
\usepackage{graphicx}  
\def\etal{{\it et\thinspace al}\ }
\def\eion{{(e~+~ion)}\ }
\begin{document}

\setcounter{page}{0}
\thispagestyle{empty}
\begin{center}
Copyright, 2008, IOP. Posted to arXiv with permission.
\resizebox{\textwidth}{!}{\includegraphics{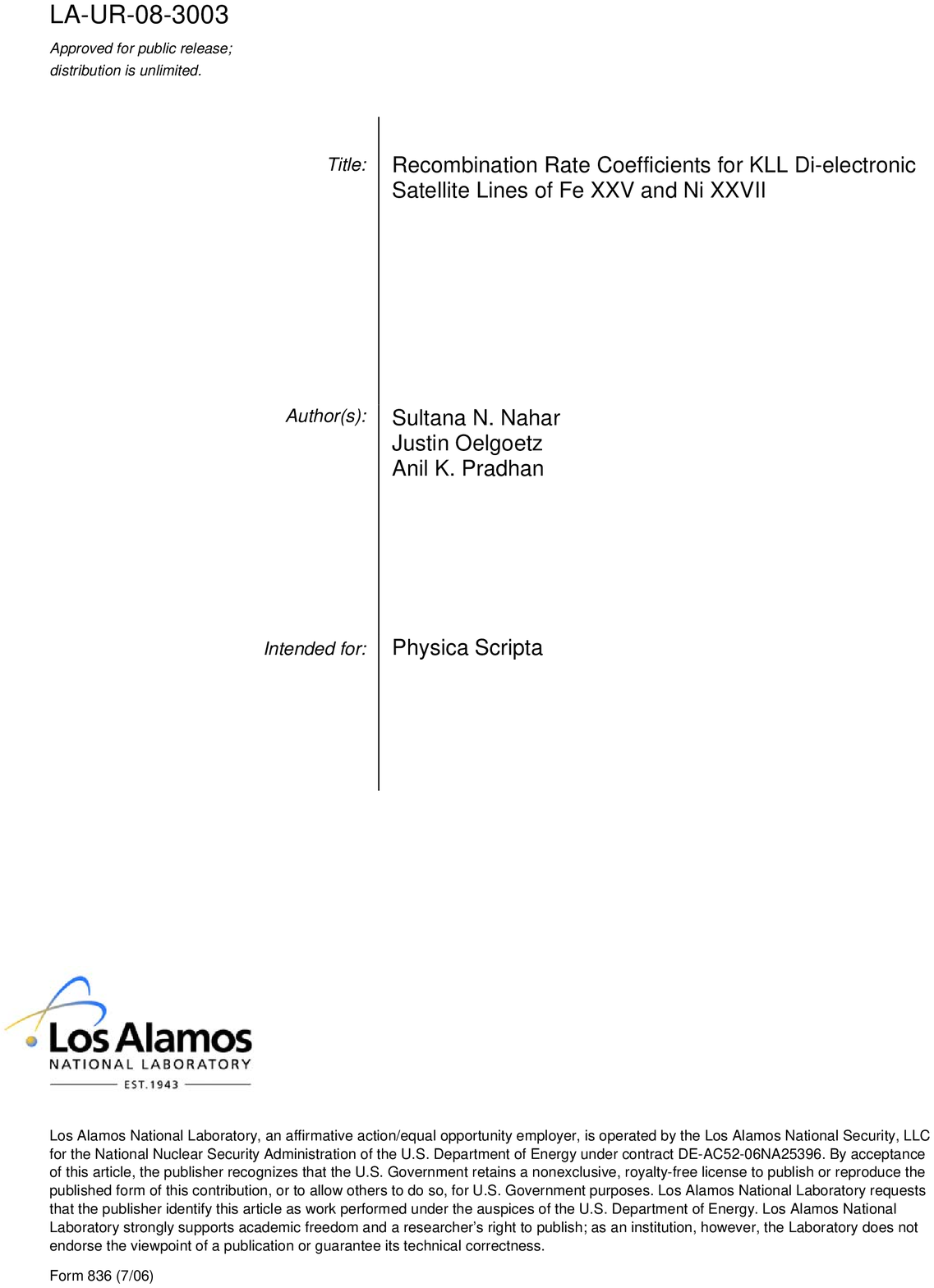}}
\end{center}
\newpage

\title[Recombination 
Rate Coefficients for KLL Di-electronic Satellite Lines]{Recombination 
Rate Coefficients for KLL Di-electronic Satellite Lines of Fe~XXV and
Ni~XXVII}

\author{S.~N. Nahar$^1$, J. Oelgoetz$^2$\footnote{Present address: Department of Physics and Astronomy, Austin Peay State University, P.O. Box 4608, Clarksville, TN 37044, USA.  oelgoetzj@apsu.edu}, A.~K. Pradhan$^1$}

\address{$^1$ Department of Astronomy, The Ohio State University,
Columbus, OH 43210, USA} 
\address{$^2$ Applied Physics Division, Los Alamos National Laboratory, Box 1336, Mail Stop F663, Los Alamos, NM 87545, USA}
\eads{\mailto{nahar@astronomy.ohio-state.edu},
\mailto{oelgoetz@lanl.gov},
\mailto{pradhan@astronomy.ohio-state.edu}}

\begin{abstract}
The unified method for total electron-ion recombination is extended to 
study the dielectronic satellite (DES) lines. These lines, formed from 
radiative decay of autoionizing states, are highly sensitive temperature 
diagnostics of astrophysical and laboratory plasma sources. The 
computation of the unified recombination rates is based on the 
relativistic Breit-Pauli R-matrix method and close coupling approximation.  
As such unified recombination cross sections ($\sigma_{RC}$) include both 
the resonant and the non-resonant background contributions and the DES 
spectra correspond directly to resonances in $\sigma_{RC}$. 
Extending the theoretical formulation developed earlier 
(Nahar S N and Pradhan A K 2006 {\it Phys.
Rev. A} {\bf 73} 062718-1) we present recombination rate
coefficients for the 
22 satellite lines of KLL complexes of helium-like Fe~XXV and Ni~XXVII. 
The isolated resonance approximation, commonly used throughout plasma 
modeling, treats these resonances essentially as bound features except 
for dielectronic capture into, and autoionization out of, 
these levels. A line profile or cross section shape
is often assumed. On the other hand, by including the coupling between the 
autoionizing and continuum channels, the unified method gives the {\it 
intrinsic} spectrum of DES lines which includes not only the energies and 
strengths, but also the natural line or cross section shapes.
A formulation is presented to derive autoionization
rates from unified resonance strengths and enable correspondence with
the isolated resonance approximation. While the rates compare very well 
with existing rates for the 
strong lines to $<$20\%, the differences for weaker DES lines
are larger. We also illustrate the application of the present results to 
the analysis of K$\alpha$ complexes observed in high-temperature X-ray 
emission spectra of Fe~XXV and Ni~XXVII. There are considerable 
differences with previous results in the total KLL
intensity for Fe~XXV at temperatures below the temperature of maximum abundance 
in coronal equilibrium.
\end{abstract}

\maketitle

\section{Introduction}

The dielectronic satellite lines are commonly seen in the spectra of 
highly ionized ions in high-temperature sources, such as active and 
flaring regions of the solar corona or fusion devices 
(Kato \etal 1998, Bely-Dubau \etal 
1982). They have been shown to be of particular relevance as diagnostics 
of both steady-state and transient stages of a plasma source (Oelgoetz 
and Pradhan 2001, 2004).
The physical process entails the collision of an electron with an 
N-electron ion core, forming an (N+1)-electron quasi-bound, 
doubly-excited resonant state, which may decay radiatively or via
auotionization (AI). Radiative decay leads to recombination of the 
free electron into a stabilised bound state of the (N+1)-electron ion
--- the process known as dielectronic recombination (DR). 
For highly charged ions of relatively heavy 
elements, the large radiative decay rates in the parent ion core 
compete effectively with the 
autoionization rates for the \eion system. 
Schematically, this branching process may be 
written as
\begin{equation}
e + X^{++} \leftrightarrow (X^+)^{**}  \leftrightarrow  \left\{
\begin{array}{c} \ e + X^{++}~(AI) \\ \  h\nu + X^+~(DR)\;. \end{array}
\right.\label{3stepproc}
\end{equation}
The doubly-excited, autoionizing state (denoted by a double asterisk) 
appears as a resonance that breaks up either via autoionization, a 
radiation-less transition back to the N-electron ion and a free electron 
carrying away the excess energy into the continuum, or via radiative 
stabilization to a recombined (N+1)-electron bound state. In the latter 
case the excited ion core decays radiatively via a strong dipole 
transition, usually into a low-lying state, and the incident electron is 
thereby captured into an (N+1)-electron state; the resulting photon 
emission manifests itself as a `dielectronic satellite' (DES) line. The 
radiative transition within 
the N-electron ion core is called a {\it principal} transition.
The DES lines are somewhat lower in energy, and at longer 
wavelengths or `redward' of, the principal line.
There is an infinite set of these 
DES lines corresponding to the Rydberg series of autoionizing levels as
$n \longrightarrow \infty$, 
converging on to the threshold for the principal core transition. Only 
the lowest $n$-complexes are usually observed and resolved however, as 
they have larger electron-capture rates (inverse of AI rates) and
sufficiently removed energetically from the principal transition. 
The higher-$n$ DES lines blend with the principal line. 

The strongest DES lines stem from the $n$=2 complex in 
He-like ions, as `satellites'
to the principal transition $1s2p$~$^1P^o_1$~$\rightarrow$~$1s^2$~$^1S_0$, 
designated as the line $w$ (also referred to as the `resonance
transition'). The  $w$-line wavelengths for Fe~XXV and Ni~XXVII are 1.8504 and
1.5884 $\AA$ respectively. The recombining electron forms a 
three-electron Li-like system in a quasi-bound state that may decay 
radiatively. The doubly-excited autoionizing states corresponding to 
the $n = 2$ complex (1s2$l$2$l'$) are designated as KLL resonances.
Radiative decay of the KLL resonances to Li-like bound states 
($1s^22s$~$^2S_{1/2}$, $1s^22p$~$^2P^o_{1/2}$, and 
$1s^22p$~$^2P^o_{3/2}$) give rise to 22 DES lines. The KLL DES lines 
from several elements are seen in high-temperature plasma sources above 
1 MK (Kato \etal 1998, Beieresdorfer \etal 1992). In order for the DES 
lines to form, the plasma temperature must be sufficiently high such that 
energetic electrons are available to recombine with the target ion via 
these quasi-bound states. A DES line arises from electrons in a 
very narrow energy range corresponding to a particular AI level and its
width. But the principal line can be 
excited by any electron with an energy greater than the threshold energy 
of excitation of the core transition. In other words, the DES line is
basically insensitive to over free-electron distribution, generally a
maxwellian, whereas the excitation of the principal line depends largely
on the electron temperature. Therefore, the ratio of a DES line to the 
principal line in the core (the aforementioned $w$-line in the case of KLL 
resonances) is very sensitive to the plasma temperature.
It follows that DES lines weaken much faster than 
the principal line near and beyond the temperature of maximum abundance 
(assuming coronal equilibrium).

The formation of dielectronic satellites may be also viewed as a subset 
of the electron-ion recombination process. As Eq. (1) implies, by taking 
detailed balance into account, \eion recombination is the inverse of 
photoionization. Recombination is usually divided into multiple components: 
radiative recombination (RR) which refers to non-resonant or background 
photo-recombination, dielectronic recombination (DR) via autoionizing 
resonances, and, if appropriate to the plasma, the stimulated analogues 
of these processes. Previous works (e.g. Nahar and Pradhan 1994, 2004) 
have developed a method that unifies both the RR and DR processes, 
accounting for quantum mechanical interference between the two processes.  

Nahar and Pradhan (2006) further extended the unified method for total 
electron-ion recombination for dielectronic satellite lines and showed 
that the profiles and intensities of the unified recombination cross 
sections ($\sigma_{RC}$) directly correspond to, and compare well with, 
the measured strengths and observed recombination spectra. We refer to 
the unified \eion recombination cross section as $\sigma_{RC}$, which is 
the detailed balance inverse of photoionization from all bound states 
of the \eion system and subsumes both the RR and DR processes. The 
unified \eion recombination method yields a representation of 
dielectronic satellite spectra that includes the interference between RR
ad DR in an {\it ab 
initio} manner. This is in contrast to existing isolated resonance 
approximation (IRA) treatments (such as Bely-Dubau \etal 1982) that do 
not include this coupling. 

In this report we extend and complete the new approach for DES lines
described in Nahar and Pradhan (2006), and demonstrate its use with 
two important practical examples. We also compute the 
resonance strengths
for KLL DES lines of (e+Fe~XXV) and (e+Ni~XXVII) to higher precision than in
the earlier work. They are presented two different ways: The
theoretical recombination resonance strength, and the DES
line strength for comparison with experimentally measured values. 
The theoretical resonance strength is computed from the unified
recombination cross sections without approximation and exhibits the
full spread of the resonance profile with energy. But the experimentally 
measured value
is obtained to one single (peak) energy; therefore, the DES line strength is
recomputed at that energy. In general, this uncertainty in the measured
values is not
too critical to accuracy for DES resonances where the peak is sharply
defined.  
 Finally, the rates for all DES lines are tabulated and should be more accurate
than previous works for spectral analysis,
potentially applicable to the analysis of X-ray observations from a 
variety of plasma sources.

\section{Theory}

As mentioned above, the unified recombination approach for the strengths 
and rates of dielectronic satellite lines is fundamentally different 
from existing approaches based on the IRA. In the IRA a DES line is 
formed when a target ion captures a free electron into an autoionizing 
state, and this autoionizing state undergoes radiative stabilization via 
decay to bound (e + ion) state to complete the dielectronic recombination 
process. This emitted photon is treated just as if it were a bound-bound 
emission line, thus it is calculated using a standard line profile, 
centered about a single energy. The unified method, on the other hand,
yields the entire line profile in an {\it ab initio} manner by calculating 
recombination cross sections from photoionization cross sections that 
account for the coupling between all of the quasi-bound states and the 
continuum, and the interference effects between the two, as summarized 
below.

\subsection{The unified method for total recombination cross sections}

The basic outline of the calculation starts with the bound N-electron
wave functions for a number of states of the target ion ($\chi_i(ion)$).
In the coupled channel R-matrix methodology, these N-electron wave 
functions are then used to calculate wave functions for the (N+1) electron 
system ($\Psi(ion+e;E,J,\pi)$, which are defined by an energy $E$, 
angular momenta $J$, and parity $\pi$) by using the close coupling 
expansion 

\begin{equation}
\Psi(ion+e;E,J,\pi) = \hat{A}\left( \sum_{i} \chi_{i}(ion)\theta_{i} + \sum_{j} c_{j} \Phi_{j}(ion+e)\right)\;,\label{ccexp}
\end{equation}

where $\chi_{i}$ is the target wave function in a specific level
$J_i\pi_i$ and $\theta_{i}$ is the wave function for the ($N$+1)-th
electron in a channel labeled as
$S_iL_i(J_i)\pi_ik_{i}^{2}\ell_i(\ J\pi))$ where $k_{i}^{2}$ is its
incident kinetic energy. A channel is open or closed depending on
positive or negative energy of the electron.  The $\Phi_j$'s are 
correlation functions of the ($N$+1)-electron system that are built 
from the one-electron wave functions of the N electron system and account 
for short range correlations and orthogonality between continuum and 
bound orbitals. It should be noted that in this treatment, autoionizing 
states are not discrete states, but instead are embedded in the continuum 
and result from the coupling of open and closed channels. As a 
consequence the terms autoionizing or quasi-bound states appear
naturally with the contiunnum and are treated together
in the unified framework; thus we will use them only to refer to the IRA
calculations.  

As the coupled channel expansion, Eq. (1), allows for interaction between the
autoionizing states and the continuum, it accounts for quantum
mechanical interference between them.  Resonances are a consequence of
this coupling. The unified
method thereby
includes both recombination processes, via the continuum (RR), 
and via autoionizing resonances (DR). In addition, it
should also be noted that the same expansion, Eq. (1), is used
for both the bound
states ($E<0$), and for the contiunuum wave functions ($E>0$).

Photoionization cross section can be obtained as
\begin{eqnarray}
\sigma_{PI}(B,E) = \sum_{(j,\pi)} \sigma_{PI}(B\rightarrow J,\pi,E)\;, 
\label{levspecific}\\
\sigma_{PI}(B\rightarrow J,\pi,E) = {1\over g_i}{4\pi^2\over 3c}\omega{\bf S(J,\pi,E)}\;,\label{sigPI}\\
{\bf S}=\left|\left<\Psi_B \left|\left| {\bf D} \right|\right| \Psi_F(J,\pi,E) \right>\right|^2\;,\label{smat}
\end{eqnarray}
where, $g_i$ is the statistical weight factor of the initial bound state, 
$\Psi_B$ and $\Psi_{F}$ respectively are the bound and free electron wave 
functions calculated using the expansion presented in (\ref{ccexp}), and 
${\bf D}$ is the dipole operator (${\bf D}= \sum\limits_i^{N+1}{r_i}$).

The radiative decay rates of highly charged H- and He-like recombining 
ions are often comparable to the Auger rates of autoionizing states, typically 
10$^{12}$ - 10$^{14}~sec^{-1}$ (e.g.  Nahar \etal  2000) with strong 
dipole allowed $2p$~($^2P^o$)~$\rightarrow$~$1s$~($^2S$) and 
$1s2p$~($^1P^o_1$)~$\rightarrow$~$1s^2$~($^1S_0$) transitions. Because of 
this it is essential to use a radiative damping treatment that will 
account for the probability of the absorbed photon being re-emitted via 
the radiative decay of the final \eion state. These calculations include 
radiative damping by using the resonance fitting procedure outlined in 
Sakimoto \etal (1990), Pradhan and Zhang (1997), and Zhang \etal (1999).

The photo-recombination cross section, $\sigma_{\rm RC}$, is related
to photoionization cross section, $\sigma_{\rm PI}$, through principle
of detailed balance as
\begin{equation}
\sigma_{\rm RC}(B,\epsilon) =
{\alpha^2 \over 4} {g_i\over g_j}{(\epsilon + I)^2\over \epsilon}
\sigma_{\rm PI}(B,\epsilon+I)\;,\label{sigRC}
\end{equation}
\noindent
in Rydberg units. Here, $\alpha$ is the fine structure constant, $\epsilon$
is the photoelectron energy, $g_j$ is the statistical weight of the 
recombined ion, and $I$ is the ionization potential ($E=\epsilon+I$). The 
recombination cross section, $\sigma_{\rm RC}$, is calculated at a 
sufficiently large number of energies to delineate the non-resonant 
background and the resonances (which, when taken together, subsume both 
radiative and dielectronic recombination processes).  It should be noted 
that (\ref{sigRC}) is valid for both $\sigma_{PI}(B,E)$ and 
$\sigma_{PI}(B\rightarrow J,\pi,E)$.

We assume that the recombining ion is in the ground state and
recombination can take place into the ground or any of the excited
recombined (e+ion) states. Thus we can calculate a total unified
$\sigma_{\rm RC}$ by summing over the final recombined states.  It
should be noted that a single resonance in the total $\sigma_{\rm RC}$
might appear in multiple level specific $\sigma_{\rm RC}$'s.  This is
because one free wave function (or autoionizing state in the IRA
framework) can often recombine into multiple bound states.  (The
selection rules for this recombination are the same as the dipole
operator, namely: $\Delta\pi=\pm 1$, $\Delta J=\pm1,0$ except $\Delta
J\ne0$ for $J=0$.)  As these final states have different energies,
different DES emission lines are formed from each pathway. There is no
interference between lines arising from the multiple recombination
pathways, although the lines do blend in the total recombination cross
section because the total cross section is simply a sum of the level
specific cross sections.  This should not be confused with interference
between RR and DR, as well as the interference between different
resonances within the same symmetry ($J\pi$).  These two interference
effects are included as a consequence of the close coupling expansion
employed in the unified method.

\subsection{Resonance strengths, rate coefficients, and intensities of
DES lines}

The resonances in unified recombination cross sections ($\sigma_{RC}$) 
are observed in the emission spectra as DES lines. As explained in Nahar 
and Pradhan (2006), integration or averaging over the resonance profiles 
provides the (a) recombination resonance strengths which are {\it intrinsic}
quantities independent of external plasma conditions, and (b) DES
intensities which depend mainly on the electron temperature. Following 
the notation of Nahar and Pradhan (2006), the recombination rate through a 
satellite line, for an electron temperature $T$, ($\alpha_s(T)$) can be 
expressed as the product of a temperature dependent term ($f(T)$) and a 
term that is intrinsic to the satellite line ($S_{\rm RC}(s)$).  Nahar 
and Pradhan (2006) define these as
\begin{eqnarray}
\alpha_s(T) = f(T)S_{\rm RC}(s)\;,\label{NP2006a}\\
f(T)={4\over \sqrt{2\pi m_e}}{e^{-{\epsilon_s\over kT}}\over{(kT)^{3/2}}}\;,\label{NP2006b}\\
S_{\rm RC}(s)=\int_{E_i}^{E_f}{\epsilon~\sigma_{RC}(\epsilon) d\epsilon}\;,\label{NP2006c}
\end{eqnarray}
where $E_i$ and $E_f$ are the upper and lower bounds for the line $s$, $m_e$ 
is the mass of the electron, $k$ is the Boltzmann constant, and $\epsilon_s$ 
is the resonance's peak energy. The temperature independent part, 
$S_{\rm RC}$, is referred to as the {\it recombination resonance strength}.

As the lines are narrow, and the resonance profiles are fairly symmetric, 
the computed DES resonance strengths can be compared directly with those 
derived from measurements of the {\it satellite line strength} ($S(s)$), 
which is related to the recombination resonance strength ($S_{\rm RC}(s)$) 
approximately as
\begin{equation}
S(s) \approx S_{\rm RC}(s)/\epsilon_s\;.\label{S}
\end{equation}
A more convenient expression relating the basic quantities above is
\begin{equation}
\alpha_s(T)=S(s)\epsilon_sf(T)=0.015484~{\epsilon_s e^{-{\epsilon_s\over
kT}}\over {T}^{3/2}} S(s)\;,\label{convenient}
\end{equation}
which is valid for any satellite line with a narrow energy width. The 
units employed are: $\epsilon$ in Rydbergs (Ry) and $\sigma_{RC}$ in 
Megabarns (Mb). The DES strength $S_{\rm RC}(s)$ can be expressed in CGS 
units via the following conversion: Ry$^2$Mb$=4.75109\times 10^{-40}$ 
ergs$^2$~cm$^2$.

One disadvantage inherent in the recombination cross section 
$\sigma_{\rm RC}$ is that it may not be used over the entire energy range 
since it diverges at zero photoelectron energy. Therefore, we also 
compute the recombination collision strength  $\Omega_{RC}$ which shows 
no such divergence since
\begin{equation}
\sigma_{RC} (Mb) = {\pi\over g_ik_i^2} \Omega_{RC}
(a_0^2/10^{-18})\;,\label{omega}
\end{equation}
where $k_i^2$ is the energy of the incident electron (equivalent to the
photoelectron energy $\epsilon$). When calculating DES rate coefficients, 
$\alpha_s(T)$, we employ both the $\sigma_{RC}$ and $\Omega_{RC}$ as a 
numerical consistency check.

Intensities of the individual DES lines can be obtained as,
\begin{equation}
I_s(i\rightarrow j,T)=\alpha_s(T) n_i n_e\;,\label{intensity}
\end{equation}
where $n_i$ is the density of the target ion and $n_e$ is the electron
density. However, it is common, and often more useful when comparing to 
experiment, to compare the intensity ratio of the satellite line to the 
dipole core excitation line. In the case of KLL lines, this is the
$w$-line due to the transition
$1s2p$~~$^1P^o_1$~$\rightarrow$~$1s^2$~~$^1S_0$.
The intensity 
ratio of a KLL satellite line to the $w$-line is obtained as
\begin{equation}
\frac{I_s}{I_w}=\frac{\alpha_s}{q_w}\;,\label{Iratio}
\end{equation}
where $q_w$ is the rate coefficient for collisional excitation from the 
ground state, into the $1s2p$~($^1P^o_1$) state which gives rise to the 
w line.

\subsection{Autoionization rates from unified satellite strengths}

Most of the modeling codes are based on the IRA. They calculate
dielectronic capture and recombination rates 
from autoionization rates $A_a$ and radiative rates $A_r$. 
Hence, it is desirable to establish a 
correspondence between the present and the earlier approaches by 
formulating the autoionization rates ($A_a$) from unified resonance 
strengths. This will facilitate the use of the present data in modeling 
codes that are based on the IRA framework.

Dielectronic recombination, in the IRA is a multi-step process, the first 
step of which is to calculate a dielectronic capture rate ($D$) which is 
related to the autoionization rate by
\begin{equation}
D_{m \rightarrow i} = \frac{g_i}{g_m}\frac{h^3e^{\frac{-\epsilon_s}{kT}}}{2(2\pi m_ekT)^{
\frac{3}{2}}}A_{a}(i \rightarrow m)\;,
\label{eqnDC}
\end{equation}
where $m$ is the state of the target ion, and $i$ refers to the 
autoionizing state. It should be noted that this result assumes that 
dielectronic capture is the detailed balance inverse of autoionization, 
but does not assume a specific line (resonance) shape, other than that the 
shape is centered about $\epsilon_s$. The total DR rate coefficient for
the 
satellite line is then simply the product of the dielectronic capture rate 
and the branching ratio,
\begin{equation}
\alpha^{DR}_s(T)=D_{m \rightarrow i}\frac{A_r(i\rightarrow j)}
{\sum\limits_lA_r(i \rightarrow l)+\sum\limits_kA_a(i\rightarrow k)}\;,
\label{eqndefira-s}
\end{equation}
where an ion in state $m$ captures an electron to form autoionizing state 
$i$, which then completes the DR process by radiatively stabilizing 
(decaying) to state $j$. The $A_r$'s are radiative decay rates for this 
last step. The present approach equates this two step process to a 
resonance in the level specific recombination cross section 
($\sigma_{\rm RC}$). The resonances in $\sigma_{\rm RC}$ correspond to 
complete dielectronic satellite line profiles. Substituting (\ref{eqnDC}) 
into (\ref{eqndefira-s}) and approximating $\alpha^{DR}_s(T)$ by 
$\alpha_s(T)$, one gets
\begin{equation}
4S_{\rm RC}=\frac{g_i}{g_m}\frac{h^3}{4\pi m_e}A_a(i \rightarrow n)\frac{A_r(i
\rightarrow j)}{\sum\limits_lA_r(i \rightarrow l)+\sum\limits_kA_a(i
\rightarrow k)}\;, \label{eqnsub}
\end{equation}
which gives the autoionization rate $A_a(i\rightarrow m)$,
\begin{equation}
\fl A_a(i \rightarrow m) = \frac{S_{\rm RC}}{\frac{g_i}{g_m}\frac{h^3}{16
\pi m_e}A_r(i \rightarrow j) - S_{\rm RC}} \times \left(
\sum\limits_lA_r(i \rightarrow l) + \sum\limits_{k\ne m} A_a(i
\rightarrow k)\right)\;.\label{bigAasat}
\end{equation}

The above equation shows that a given autoionization rate depends on the 
degeneracy factors, some constants, $S_{\rm RC}(s)$, the radiative decay 
rates out of the state, and the other autoionization rates. Hence if the 
radiative decay rates are available from another source, $A_a$ for other 
continuum states can be obtained by solving the set of coupled linear 
equations arise from (\ref{bigAasat}) provided the $S_{\rm RC}$ value for 
the resonances are known. In the case of KLL lines however:
$\sum\limits_{k\ne m} A_a(i \rightarrow k) = 0$ and the expression
simplifies to,
\begin{equation}
A_a(i \rightarrow m)=\frac{S_{\rm RC}}{\frac{g_i}{g_m}\frac{h^3}{16\pi
m_e}A_r(i \rightarrow j) - S_{\rm RC}} \sum\limits_lA_r(i \rightarrow l)\;.\label{Aasimple}
\end{equation}

It should also be noted that this approximation assumes that the RR 
background in the unified recombination cross section ($\sigma_{\rm RC}$) 
has a negligible contribution to the value of $S_{\rm RC}(s)$ for each and 
every $s$. In general, this is a reasonable approximation for the systems 
under consideration in this work.

\subsection{Identification and resonance strengths of DES lines}

Proper and complete spectroscopic identification of the DES lines with 
respect to the energy positions is not straightforward from the unified 
recombination cross sections. The Breit-Pauli R-matrix (BPRM) calculations 
do not identify resonances spectroscopically a priori, as they are formed 
through coupling or interference of open and closed channels. They can be 
identified easily from the energy positions of the resonances which are 
known for ions, such as Fe~XXV, from existing experimental values. They 
can also be determined theoretically from relevant spectroscopic 
transitions in atomic structure calculations. In our approach, we sort out 
the DES lines from the level-specific recombination cross sections 
$\sigma_{RC}(nSLJ)$ by matching the resonances in the cross sections with 
both the resonances in the total cross section (see Nahar and Pradhan 2006) 
and the values given in Gabriel (1972).

Again it should be noted that a single resonance in the total 
recombination cross section may be the sum of individual resonances that 
appear at the same energy in multiple level specific cross sections.  
This is because the same free wave function (or autoionizing state in the 
IRA framework) can recombine into multiple final states. These resonances, 
while they do not quantum mechanically interfere, blend together in the 
total cross section. In such cases $S_{\rm RC}(s)$ must be calculated from 
the level specific cross sections.

The fractional contribution of the DES line to the resonance in the total 
cross section can be determined by examining the contributions of all the
constituent lines. The fractions $x_s$ are determined from the ratio of 
the line strengths, $S_{\rm RC}(s)$, calculated from the level specific
recombination cross sections to the summed strengths 
$\sum\limits_jS_{\rm RC}(j)$, that is, $x_s=S_{\rm RC}(s)/
\sum\limits_jS_{\rm RC}(s)$.

\section{Computation}

Computation for the present approach involved extension of existing 
codes and developing a new code to compute satellite resonance strengths 
using unified recombination cross sections. The recombination rates of 
all satellite lines were obtained in four separate computations; the 
purpose of the redundancy is to test for numerical problems and 
consistencies.

Calculations for photoionization and electron-ion recombination are
done in several stages, starting with atomic structure calculations 
using the code SUPERSTRUCTURE (Eissner \etal 1974) for the target ion
wave function. The one-electron orbitals of the target are the initial 
input to the BPRM suites of codes (Berrington \etal 1987, 1995, which were 
developed under the Iron Project (IP), see: Hummer \etal 1993). The 
total wave function expansion for the resonance states of Li-like Fe~XXIV 
consists of a 13-level expansion of the target He-like ion, Fe~XXV, (Nahar 
\etal  2001) and a 17-level expansion for Ni~XXVII (Nahar 2005). 
Radiation damping of the resonances is included as described in (Zhang 
\etal 1999). The present computations for $\sigma_{\rm PI}$ are confined 
only to photoelectron energy regions of the first resonance complex, 
which is the KLL complex, and only to partial photoionization cross 
sections for leaving the core ion in the ground $1s^2$~($^1S_0$) level.

The level specific recombination cross sections ($\sigma_{\rm RC}(i)$),
collision strengths ($\Omega_{\rm RC}(E,i)$), and rates ($\alpha_R(E,i)$)
are computed for all coupled symmetries and levels, and summed to
obtain the total using the program PBPRRC (Nahar \etal 2000). The
computation for $\sigma_{\rm RC}(i)$ has been carried out in specified
energy ranges ($\epsilon_i-\epsilon_f$) corresponding to each
satellite line $s$ using the extended program RECXS. The total
recombination rate coefficients, $\alpha_R$, are also obtained from the
total recombination collision strength, $\Omega_{\rm RC}$. The difference
between the two numerical values is a few percent.

A new code SATLN is written to process unified $\sigma_{\rm RC}(i)$ and
obtain various quantities, such as the location and identification of 
the satellite lines, the dielectronic satellite strengths, $S_{\rm RC}(s)$ 
and $S(s)$, and the recombination rate coefficients of the individual 
and integrated satellite strengths, as well as to carry out internal 
consistency checks. The program also determines the individual satellite 
line contributions to blended resonances in the recombination spectra.

 We perform several checks on the numerical accuracy of the
calculations, as well as to
ascertain if background contributions to DES intensities are negligible.
The recombination rate coefficients were obtained in four
different ways. The first set of rates was obtained by summing the 
contributions of level specific recombination rate coefficients for each 
of the final three J$\pi$ symmetries responsible for the DES spectra. 
The second set of rates was obtained from integration of the total 
recombination collision strengths calculated directly from 
photoionization cross sections, and including background cross 
sections for recombination into high-n states (comprising the RR 
contribution). The third set of results is obtained via integration of 
total $\sigma_{\rm RC}$ for the entire energy range of the KLL complex,
multiplied by the Maxwellian temperature factor $f(t)$ (see (8)).
The fourth set of calculations is to compute the DES rates individually. 
The rate coefficients from all four separate calculations are in good 
agreement with each other, providing a numerical consistency check as 
well as validation that the background contributions due to RR are not 
significant.

\section{Results and Discussions}

Results for Fe~XXV and Ni~XXVII from the new approach of using unified 
method for positions, strengths, recombination rate coefficients and 
intensities of the DES lines are discussed in subsections below.

\subsection{Recombination Spectrum of DES lines: Profiles and Energies}

\begin{figure}
   \begin{center}
      \resizebox{120mm}{!}{\includegraphics{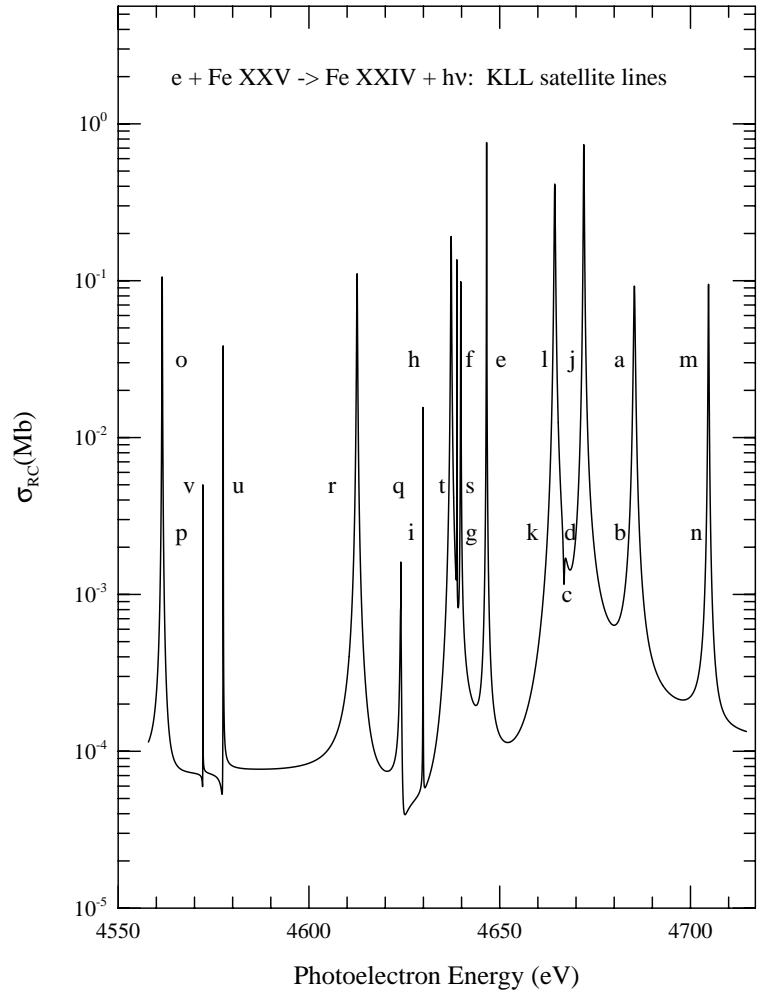}}
\caption{\label{rc-fe}Theoretical profiles of the 22 dielectronic
satellite lines of the K${\alpha}$ complex seen in unified recombination 
cross sections of Fe~XXV. The interference effects in the {\it ab
initio} 
results manifest themselves in the natural overlap of line profiles as 
described in the text.}
\end{center}
\end{figure}

\begin{figure}
   \begin{center}
      \resizebox{120mm}{!}{\includegraphics{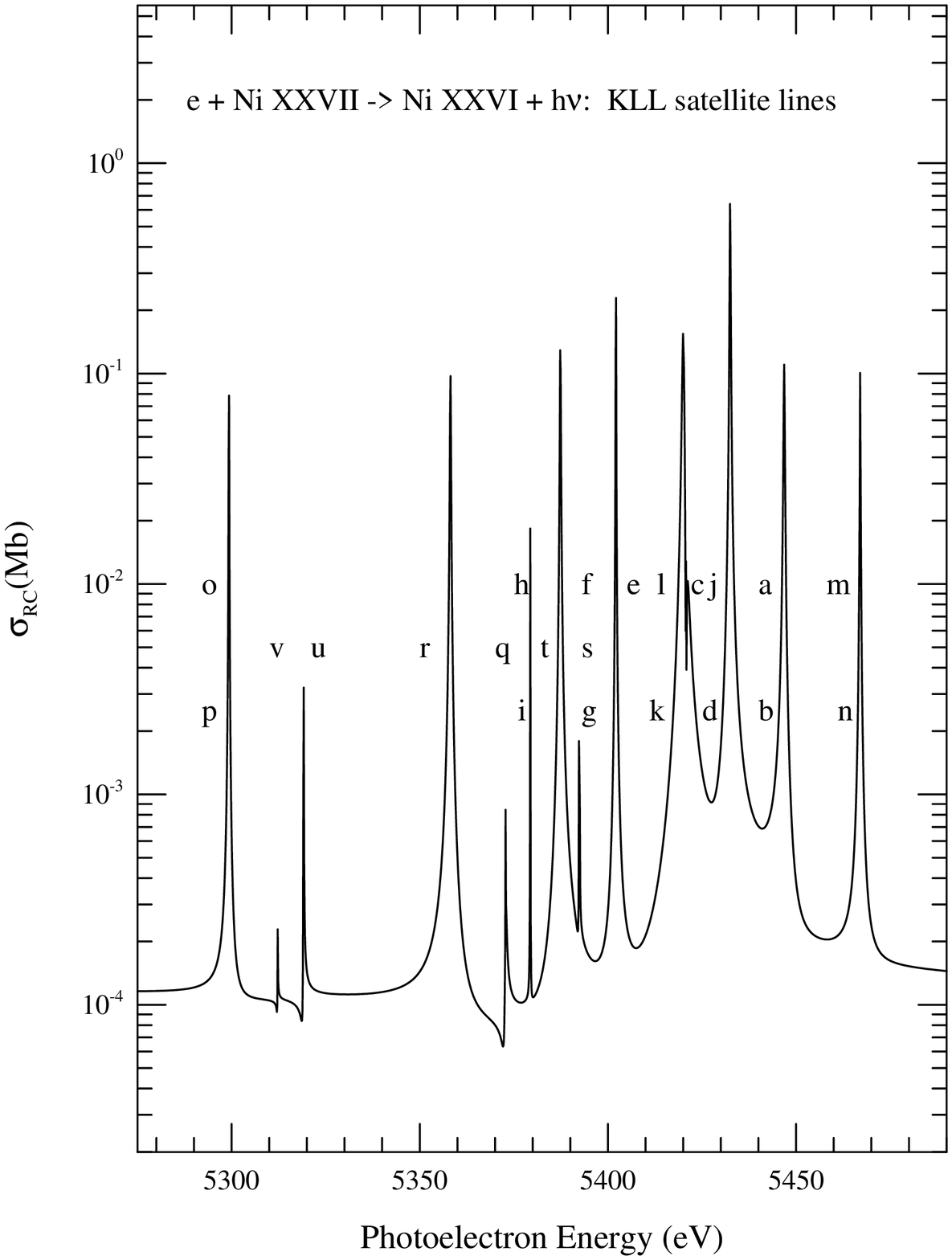}}
\caption{\label{rc-ni}Theoretical profiles of the 22 dielectronic
satellite lines of the K${\alpha}$ complex seen in unified
recombination cross sections of Ni~XXVII, as in figure 1.}
\end{center}
\end{figure}

The unified method for total electron-ion recombination can generate
the recombination spectrum of dielectronic satellite lines including
the background as shown in figure \ref{rc-fe} for Fe~XXV and figure 
\ref{rc-ni} for Ni~XXVII. 
The autoionizing resonances in the KLL complex of a He-like ion that 
appear in the total unified $\sigma_{\rm RC}$ are the DES lines and
provide the relevant physical quantities of the satellite lines. As 
mentioned previously, the resonances are formed via a process that is 
analogous to the radiative decay of the autoionizing states arising from 
the set of configurations 1s2$l$2$l'$ into the three n=2 levels: 
1s$^2$2s~$^2S_{1/2}$, 1s$^2$2p~$^2P^o_{1/2}$, and 
1s$^2$2p~$^2P^o_{3/2}$. The 22 KLL resonances or satellite lines for 
the two ions are listed in tables 1 and 2.

\begin{table}
\fl
\caption{\label{arttype} The 22 dielectronic satellite lines of KLL 
complex in recombination of (e~+~Fe~XXV): the alphabetic notation 
(following Gabriel 1972), corresponding resonant transition, 
experimentally measured energy position E$_{ex}$, computed energy 
position in the unified method E$_P$, satellite strengths S(s) in 
$10^{-20}{\rm cm}^2$eV, recombination resonance strength $S_{RC}$ in 
$10^{-40}{\rm cm}^2$erg$^2$, and fractional contribution x$_s$ to each 
resolved DES lines.}
\footnotesize\rm
\begin{tabular}{rllllllr} 
\br
\multicolumn{2}{r}{Index} & \multicolumn{1}{c}{Transition} & 
\multicolumn{1}{c}{E$_{ex}$} & \multicolumn{1}{c}{E$_P$} & 
\multicolumn{1}{c}{S(s)} & \multicolumn{1}{c}{S$_{RC}$} & 
\multicolumn{1}{c}{x$_s$}\\
\multicolumn{1}{c}{} & \multicolumn{1}{c}{} & \multicolumn{1}{c}{} & 
\multicolumn{1}{c}{eV} & \multicolumn{1}{c}{eV} & \multicolumn{1}{c}{} & 
\multicolumn{1}{c}{} & \multicolumn{1}{c}{}\\
\mr
  1& o& $1s2s^2(^2S_{1/2})\rightarrow 1s^22p(^2P^o_{3/2})$      & 4553.4& 4561.53&   0.8832&   1.034&  0.50\\
  2& p& $1s2s^2(^2S_{1/2})\rightarrow 1s^22p(^2P^o_{1/2})$      & 4553.4& 4561.54&   0.8820&   1.033&  0.50\\
  3& v& $1s2p\;^3P^o\;2s(^4P^o_{1/2})\rightarrow 1s^22s(^2S_{1/2})$ & 4573.9& 4572.19&   0.05994&   0.07033&  1.00\\
  4& u& $1s2p\;^3P^o\;2s(^4P^o_{3/2})\rightarrow 1s^22s(^2S_{1/2})$ & 4578.9& 4577.52&   0.1628&   0.1913&  1.00\\
  5& r& $1s2p\;^1P^o\;2s(^2P^o_{1/2})\rightarrow 1s^22s(^2S_{1/2})$ & 4615.1& 4612.63&   3.798&   4.495&  1.00\\
  6& q& $1s2p\;^1P^o\;2s(^2P^o_{3/2})\rightarrow 1s^22s(^2S_{1/2})$ & 4625.3& 4622.54&   0.08194&   0.09721&  1.00\\
  7& i& $1s2p^2(^4P_{1/2})\rightarrow 1s^22p(^2P^o_{1/2})$      & 4624.6& 4629.91&   0.07853&   0.09331&  0.93\\
  8& h& $1s2p^2(^4P_{1/2})\rightarrow 1s^22p(^2P^o_{3/2})$      & 4624.6& 4629.92&   0.00554&   0.00659&  0.07\\
  9& t& $1s2p\;^3P^o\;2s(^2P^o_{1/2})\rightarrow 1s^22s(^2S_{1/2})$ & 4632.9& 4635.60&   5.517&   6.564&  1.00\\
 10& g& $1s2p^2(^4P_{3/2})\rightarrow 1s^22p(^2P^o_{1/2})$      & 4639.4& 4638.77&   0.01199&   0.01427&  0.03\\
 11& f& $1s2p^2(^4P_{3/2})\rightarrow 1s^22p(^2P^o_{3/2})$      & 4632.9& 4638.77&   0.3721&   0.4430&  0.97\\
 12& s& $1s2p\;^3P^o\;2s(^2P^o_{3/2})\rightarrow 1s^22s(^2S_{1/2})$ & 4642.5& 4639.79&   1.291&   1.538&  1.00\\
 13& e& $1s2p^2(^4P_{5/2})\rightarrow 1s^22p(^2P^o_{3/2})$      & 4639.0& 4646.57&   0.4849&   5.783&  1.00\\
 14& k& $1s2p^2(^2D_{3/2})\rightarrow 1s^22p(^2P^o_{1/2})$      & 4658.1& 4664.43&   17.43&   20.85&  0.92\\
 15& l& $1s2p^2(^2D_{3/2})\rightarrow 1s^22p(^2P^o_{3/2})$      & 4658.1& 4664.43&   1.445&   1.729&  0.08\\
 16& d& $1s2p^2(^2P_{1/2})\rightarrow 1s^22p(^2P^o_{1/2})$      & 4658.6& 4667.21&   0.7809&   0.9354&  0.82\\
 17& c& $1s2p^2(^2P_{1/2})\rightarrow 1s^22p(^2P^o_{3/2})$      & 4658.6& 4666.78&   0.1703&   0.2040&  0.18\\
 18& j& $1s2p^2(^2D_{5/2})\rightarrow 1s^22p(^2P^o_{3/2})$      & 4664.1& 4672.05&   26.82&   32.18&  1.00\\
 19& a& $1s2p^2(^2P_{3/2})\rightarrow 1s^22p(^2P^o_{3/2})$      & 4677.0& 4685.28&   6.118&   7.359&  0.97\\
 20& b& $1s2p^2(^2P_{3/2})\rightarrow 1s^22p(^2P^o_{1/2})$      & 4677.0& 4685.30&   0.2116&   0.2546&  0.03\\
 21& m& $1s2p^2(^2S_{1/2})\rightarrow 1s^22p(^2P^o_{3/2})$      & 4697.7& 4705.47&   2.753&   3.326&  0.95\\
 22& n& $1s2p^2(^2S_{1/2})\rightarrow 1s^22p(^2P^o_{1/2})$      & 4697.7& 4704.73&   0.1301&   0.1571&  0.05\\
\br
\end{tabular}
\end{table}

\begin{table}
\caption{The 22 dielectronic satellite lines of KLL complex in 
recombination of (e~+~Ni~XXVI): the alphabetic notation (following 
Gabriel 1972), corresponding resonant transition, computed energy in 
the unified method E$_P$, satellite strengths S(s) in $10^{-20}{\rm cm}^2$eV, 
recombination resonance strength $S_{\rm RC}$ in $10^{-40} ~cm^2erg^2$, 
and fractional contribution x$_s$ to each resolved DES lines.}
\footnotesize\rm
\begin{tabular}{rllllllr} 
\br
\multicolumn{2}{r}{Index} & \multicolumn{1}{c}{Transition} & 
\multicolumn{1}{c}{E$_P$} & \multicolumn{1}{c}{S(s)} & 
\multicolumn{1}{c}{S$_{RC}$} & \multicolumn{1}{c}{x$_s$}\\
\multicolumn{1}{c}{} & \multicolumn{1}{c}{} & \multicolumn{1}{c}{} & 
\multicolumn{1}{c}{eV} & \multicolumn{1}{c}{eV} & \multicolumn{1}{c}{} & 
\multicolumn{1}{c}{} & \multicolumn{1}{c}{} \\
\mr
  1& o& $1s2s^2(^2S_{1/2})\rightarrow 1s^22p(^2P^o_{3/2})$      &  5299.32& 0.9864&   1.341&  0.47\\
  2& p& $1s2s^2(^2S_{1/2})\rightarrow 1s^22p(^2P^o_{1/2})$      &  5299.32& 1.115&   1.516&  0.53\\
  3& v& $1s2p\;^3P^o\;2s(^4P^o_{1/2})\rightarrow 1s^22s(^2S_{1/2})$ &  5312.25& 0.07304& 0.09958&  1.00\\
  4& u& $1s2p\;^3P^o\;2s(^4P^o_{3/2})\rightarrow 1s^22s(^2S_{1/2})$ &  5319.17& 0.1765& 0.2410&  1.00\\
  5& r& $1s2p\;^1P^o\;2s(^2P^o_{1/2})\rightarrow 1s^22s(^2S_{1/2})$ &  5358.17& 4.256&   5.852&  1.00\\
  6& q& $1s2p\;^1P^o\;2s(^2P^o_{3/2})\rightarrow 1s^22s(^2S_{1/2})$ &  5372.30& 0.1059& 0.1461&  1.00\\
  7& i& $1s2p^2(^4P_{1/2})\rightarrow 1s^22p(^2P^o_{1/2})$      &  5379.37& 0.2769& 0.3823&  0.98\\
  8& h& $1s2p^2(^4P_{1/2})\rightarrow 1s^22p(^2P^o_{3/2})$      &  5379.37& 0.00427&0.005885&  0.02\\
  9& t& $1s2p\;^3P^o\;2s(^2P^o_{1/2})\rightarrow 1s^22s(^2S_{1/2})$ &  5387.31& 5.364&   7.415&  1.00\\
 10& s& $1s2p\;^3P^o\;2s(^2P^o_{3/2})\rightarrow 1s^22s(^2S_{1/2})$ &  5392.30& 0.1284& 0.1777&  0.81\\
 11& g& $1s2p^2(^4P_{3/2})\rightarrow 1s^22p(^2P^o_{1/2})$      &  5392.41& 0.00379&0.005241&  0.02\\
 12& f& $1s2p^2(^4P_{3/2})\rightarrow 1s^22p(^2P^o_{3/2})$      &  5392.41& 0.02548&0.03.527&  0.16\\ 
 13& e& $1s2p^2(^4P_{5/2})\rightarrow 1s^22p(^2P^o_{3/2})$      &  5402.16& 5.401&   7.489&  1.00\\ 
 14& k& $1s2p^2(^2D_{3/2})\rightarrow 1s^22p(^2P^o_{1/2})$      &  5419.96& 6.380&   8.866&  0.51\\ 
 15& l& $1s2p^2(^2D_{3/2})\rightarrow 1s^22p(^2P^o_{3/2})$      &  5419.96& 6.100&   8.477&  0.49\\ 
 16& d& $1s2p^2(^2P_{1/2})\rightarrow 1s^22p(^2P^o_{1/2})$      &  5421.21& 1.161&   1.617&  0.50\\ 
 17& c& $1s2p^2(^2P_{1/2})\rightarrow 1s^22p(^2P^o_{3/2})$      &  5421.21& 1.151&   1.602&  0.50\\ 
 18& j& $1s2p^2(^2D_{5/2})\rightarrow 1s^22p(^2P^o_{3/2})$      &  5432.44& 25.04&   34.92&  1.00\\
 19& a& $1s2p^2(^2P_{3/2})\rightarrow 1s^22p(^2P^o_{3/2})$      &  5446.84& 5.993&   8.380&  0.96\\
 20& b& $1s2p^2(^2P_{3/2})\rightarrow 1s^22p(^2P^o_{1/2})$      &  5446.84& 0.2348& 0.3283&  0.04\\
 21& m& $1s2p^2(^2S_{1/2})\rightarrow 1s^22p(^2P^o_{3/2})$      &  5467.02& 2.328&   3.267&  0.78\\
 22& n& $1s2p^2(^2S_{1/2})\rightarrow 1s^22p(^2P^o_{1/2})$      &  5467.02& 0.6634& 0.9.310&  0.22\\
\br
\end{tabular}
\end{table}

The interference effects of resonant lines, not considered in IRA methods, 
manifest themselves in overlapping satellite profiles (the most prominent 
example is the interaction of d and k, or c and l). Also included in the 
total unified recombination cross sections (shown in figures \ref{rc-fe} 
and \ref{rc-ni}) are the multitude of non-interfering, yet blending lines 
(o and p, d and c, l and k, etc.). The resolved total recombination 
spectra show 15 DES lines of Fe~XXV and 14 DES lines of Ni~XXVII. The 
satellite resonances vary over orders of magnitude in cross section, with 
often overlapping profiles {\it within each symmetry}; these overlapping 
profiles give rise to interference. The pairs (k+d) and (l+c) in the 
Fe~XXV spectrum not only have overlapping profiles, but one of the 
resonances in each of the pairs is extremely weak ($<$ 1\%) and lies in 
the wings of the other one. We emphasize that the overlap and the 
resulting uncertainty are natural features, which are commonly neglected 
by IRA methods.

Owing to the natural widths of autoionizing resonances, satellite 
line profiles in the unified spectrum overlap. These overlaps can be
significant in determining satellite intensities, particularly those
of the weaker satellites.
This also implies that the satellite lines may be blended and need
to be identified carefully. 
This is in contrast to IRA methods, based on atomic structure
calculations, where level designations are initially pre-specified 
according to the
configuration, term, and level structure. As explained above, the 
lines are identified from the individual $J\pi$ contributions to the  
level-specific recombination cross sections that contain the resonances.  
Thus the KLL lines are identified from the individual $J\pi$ 
contributions to the recombination cross sections into the three Li-like 
levels $1s^22s$~$^2S_{1/2}$, $1s^22p$~$^2P^o_{1/2}$, and 
$1s^22p$~$^2P^o_{3/2}$. All 22 satellites in the $n$=2 K${\alpha}$ 
complex have been isolated and identified for both Fe~XXV and Ni~XXVII. 
The DES are labeled following the notation by Gabriel (1972).

The energies of the satellite lines are assumed to be the peaks of the 
line profiles and are given in tables 1 and 2 for Fe~XXV and Ni~XXVII 
respectively. The present theoretical energies for Fe~XXV agree well 
with previous calculations. Comparison with EBIT measurements 
(Beiersdorfer \etal 1992) shows that the experimental energies are 
systematically lower than theoretical energies by up to 8 eV ($\sim$ 
0.1\%). This may be taken as the uncertainty in the present calculations. 
Comparisons of energies and resonance strengths are presented in Nahar 
and Pradhan (2006). We know of no experimental data on the Ni~XXVII DES 
lines.

\subsection{Strengths and recombination rates of DES Lines}

The unified recombination cross sections ($\sigma_{\rm RC}$) provide the
absolute resonance strengths, $S(s)$ and $S_{RC}(s)$, of the DES lines
by direct integration. Evaluation of individual resonance strengths
may have some uncertainties due to the natural overlap of resonances.
However, overlap is not expected to be significant since the dominant
contribution to integrated value arises from a narrow energy range
around the peak energy.

The integrated resonance strengths $S(s)$ and $S_{\rm RC}(s)$ for each 
of DES lines are given in tables 1 and 2 for Fe~XXV and Ni~XXVII 
respectively. Only the strong lines in the total spectra in figures 
\ref{rc-fe} and \ref{rc-ni} provide significant rates. We have arranged 
the DES by order of increasing energy, in hopes that it might be more 
convenient in comparing with observations or experiments than the 
commonly used alphabetical order, especially where blends of DES are 
concerned.

Following satellite identifications, we obtain the fractional 
contribution, $x_s$, for each resolved line to the blended recombination 
feature (e.g.\  o and p). The $x_s$ values are given in tables 1 and 2. 
For a few weaker satellites, the interference and overlapping of 
resonances poses significant uncertainty, such as, for the weak resonances 
of the pairs (k+d) and (l+c) in the Fe~XXV spectrum.

The recombination rate coefficients ($\alpha_s(T)$) for the DES lines
are given for the recombined ion Fe~XXIV in table 3 and for Ni~XXVII in 
table 4. These are obtained from direct integration of the total 
unified recombination collision strengths $\Omega_{RC}$ as described in 
the theory section. These rate coefficients are more accurate than those using 
the approximate formula given in Nahar and Pradhan (2006), since the
full energy variation of the exponential function $e^{\epsilon_s}/kT$ is
is considered in Eq. (9),
rather the peak value of the resonance. These rates were 
checked with those obtained from sum of individual contributions of the 
level-specific rates for numerical accuracy. The fractions $x_s$ are 
used for $\alpha_s(T)$ of blended lines.
\begin{table}
\noindent
\caption{Recombination rate coefficients for dielectronic satellite lines
of Fe~XXV forming Fe~XXIV (notation: x-y $\rightarrow x\times 10^{-y}$).}
\scriptsize
\begin{tabular}{llllllllllll}
\br
\multicolumn{1}{c}{T(K)}   & \multicolumn{11}{c}{$\alpha_R(cm^3/s)$} \\
 & \multicolumn{1}{c}{o} & \multicolumn{1}{c}{p} & \multicolumn{1}{c}{v} & 
\multicolumn{1}{c}{u} & \multicolumn{1}{c}{r} & \multicolumn{1}{c}{q} & 
\multicolumn{1}{c}{i} & \multicolumn{1}{c}{h} & \multicolumn{1}{c}{t} & 
\multicolumn{1}{c}{g} & \multicolumn{1}{c}{f}\\
\mr
6.0&3.46-35&3.46-35&2.09-36&5.30-36&8.33-35&1.59-36&1.41-36&9.96-38&9.12-35&1.94-37&6.03-36\\
6.1&1.31-30&1.31-30&8.10-32&2.09-31&3.56-30&6.99-32&6.29-32&4.44-33&4.14-30&8.85-33&2.75-31\\
6.2&5.28-27&5.28-27&3.33-28&8.68-28&1.58-26&3.17-28&2.89-28&2.04-29&1.93-26&4.13-29&1.28-27\\
6.3&3.60-24&3.60-24&2.30-25&6.06-25&1.16-23&2.37-25&2.18-25&1.54-26&1.47-23&3.16-26&9.81-25\\
6.4&5.97-22&5.96-22&3.87-23&1.02-22&2.05-21&4.22-23&3.92-23&2.77-24&2.67-21&5.75-24&1.79-22\\
6.5&3.22-20&3.22-20&2.11-21&5.62-21&1.16-19&2.42-21&2.26-21&1.59-22&1.55-19&3.34-22&1.04-20\\
6.6&7.14-19&7.12-19&4.70-20&1.26-19&2.67-18&5.61-20&5.26-20&3.72-21&3.63-18&7.84-21&2.43-19\\
6.7&7.78-18&7.77-18&5.16-19&1.39-18&3.01-17&6.35-19&5.98-19&4.22-20&4.15-17&8.96-20&2.78-18\\
6.8&4.84-17&4.83-17&3.22-18&8.68-18&1.91-16&4.06-18&3.84-18&2.71-19&2.67-16&5.77-19&1.79-17\\
6.9&1.92-16&1.92-16&1.29-17&3.47-17&7.76-16&1.65-17&1.57-17&1.11-18&1.09-15&2.36-18&7.34-17\\
7.0&5.36-16&5.35-16&3.59-17&9.73-17&2.20-15&4.69-17&4.46-17&3.15-18&3.12-15&6.74-18&2.09-16\\
7.1&1.13-15&1.13-15&7.57-17&2.05-16&4.68-15&1.00-16&9.53-17&6.73-18&6.67-15&1.44-17&4.48-16\\
7.2&1.89-15&1.89-15&1.28-16&3.46-16&7.93-15&1.70-16&1.62-16&1.15-17&1.14-14&2.46-17&7.65-16\\
7.3&2.67-15&2.66-15&1.80-16&4.89-16&1.13-14&2.42-16&2.31-16&1.63-17&1.62-14&3.51-17&1.09-15\\
7.4&3.26-15&3.25-15&2.20-16&5.98-16&1.38-14&2.98-16&2.84-16&2.01-17&2.00-14&4.33-17&1.34-15\\
7.5&3.56-15&3.55-15&2.40-16&6.54-16&1.52-14&3.27-16&3.12-16&2.20-17&2.19-14&4.76-17&1.48-15\\
7.6&3.55-15&3.55-15&2.40-16&6.54-16&1.52-14&3.28-16&3.13-16&2.21-17&2.21-14&4.78-17&1.48-15\\
7.7&3.31-15&3.30-15&2.24-16&6.09-16&1.42-14&3.06-16&2.93-16&2.07-17&2.06-14&4.47-17&1.39-15\\
7.8&2.91-15&2.90-15&1.97-16&5.36-16&1.25-14&2.70-16&2.59-16&1.83-17&1.82-14&3.95-17&1.22-15\\
7.9&2.45-15&2.44-15&1.66-16&4.52-16&1.06-14&2.28-16&2.18-16&1.54-17&1.54-14&3.33-17&1.03-15\\
8.0&1.99-15&1.98-15&1.35-16&3.67-16&8.58-15&1.85-16&1.77-16&1.25-17&1.25-14&2.71-17&8.41-16\\
8.1&1.57-15&1.57-15&1.06-16&2.90-16&6.78-15&1.47-16&1.40-16&9.91-18&9.88-15&2.14-17&6.65-16\\
8.2&1.21-15&1.21-15&8.20-17&2.24-16&5.24-15&1.13-16&1.08-16&7.66-18&7.64-15&1.66-17&5.14-16\\
8.3&9.18-16&9.16-16&6.22-17&1.70-16&3.98-15&8.60-17&8.23-17&5.81-18&5.80-15&1.26-17&3.90-16\\
8.4&6.86-16&6.85-16&4.65-17&1.27-16&2.98-15&6.43-17&6.16-17&4.35-18&4.34-15&9.41-18&2.92-16\\
8.5&5.07-16&5.07-16&3.44-17&9.38-17&2.20-15&4.76-17&4.56-17&3.22-18&3.21-15&6.96-18&2.16-16\\
8.6&3.72-16&3.71-16&2.52-17&6.87-17&1.61-15&3.49-17&3.34-17&2.36-18&2.35-15&5.10-18&1.58-16\\
8.7&2.70-16&2.70-16&1.83-17&5.00-17&1.17-15&2.54-17&2.43-17&1.72-18&1.71-15&3.71-18&1.15-16\\
8.8&1.96-16&1.95-16&1.33-17&3.62-17&8.50-16&1.84-17&1.76-17&1.24-18&1.24-15&2.69-18&8.35-17\\
8.9&1.41-16&1.41-16&9.56-18&2.61-17&6.12-16&1.32-17&1.27-17&8.95-19&8.93-16&1.94-18&6.01-17\\
9.0&1.01-16&1.01-16&6.86-18&1.87-17&4.40-16&9.50-18&9.10-18&6.42-19&6.41-16&1.39-18&4.32-17\\
\mr
 & \multicolumn{1}{c}{s}& \multicolumn{1}{c}{e}& \multicolumn{1}{c}{k}&
\multicolumn{1}{c}{l}& \multicolumn{1}{c}{d}& \multicolumn{1}{c}{c}&
\multicolumn{1}{c}{j}& \multicolumn{1}{c}{a}& \multicolumn{1}{c}{b}&
\multicolumn{1}{c}{m}& \multicolumn{1}{c}{n} \\
\mr
6.0&2.07-35&7.22-35&2.12-34&1.75-35&9.24-36&2.02-36&2.99-34&5.86-35&2.03-36&2.11-35&9.98-37\\
6.1&9.46-31&3.35-30&1.02-29&8.49-31&4.50-31&9.81-32&1.47-29&2.98-30&1.03-31&1.13-30&5.32-32\\
6.2&4.43-27&1.59-26&5.02-26&4.16-27&2.21-27&4.83-28&7.32-26&1.52-26&5.26-28&5.96-27&2.81-28\\
6.3&3.39-24&1.23-23&3.99-23&3.31-24&1.77-24&3.85-25&5.89-23&1.25-23&4.31-25&5.03-24&2.38-25\\
6.4&6.18-22&2.25-21&7.48-21&6.21-22&3.32-22&7.24-23&1.11-20&2.40-21&8.30-23&9.91-22&4.68-23\\
6.5&3.59-20&1.32-19&4.46-19&3.69-20&1.98-20&4.32-21&6.69-19&1.46-19&5.04-21&6.13-20&2.90-21\\
6.6&8.44-19&3.12-18&1.07-17&8.84-19&4.75-19&1.04-19&1.61-17&3.54-18&1.22-19&1.51-18&7.14-20\\
6.7&9.65-18&3.58-17&1.24-16&1.03-17&5.51-18&1.20-18&1.87-16&4.16-17&1.44-18&1.80-17&8.49-19\\
6.8&6.23-17&2.31-16&8.07-16&6.69-17&3.60-17&7.85-18&1.23-15&2.74-16&9.48-18&1.20-16&5.65-18\\
6.9&2.55-16&9.50-16&3.33-15&2.76-16&1.49-16&3.25-17&5.09-15&1.14-15&3.95-17&5.02-16&2.37-17\\
7.0&7.28-16&2.72-15&9.59-15&7.95-16&4.28-16&9.34-17&1.47-14&3.30-15&1.14-16&1.46-15&6.90-17\\
7.1&1.56-15&5.83-15&2.07-14&1.71-15&9.24-16&2.01-16&3.17-14&7.16-15&2.48-16&3.18-15&1.50-16\\
7.2&2.66-15&9.96-15&3.54-14&2.94-15&1.58-15&3.46-16&5.44-14&1.23-14&4.26-16&5.49-15&2.59-16\\
7.3&3.79-15&1.42-14&5.06-14&4.20-15&2.26-15&4.94-16&7.78-14&1.77-14&6.11-16&7.89-15&3.73-16\\
7.4&4.67-15&1.75-14&6.26-14&5.19-15&2.80-15&6.11-16&9.63-14&2.19-14&7.57-16&9.81-15&4.63-16\\
7.5&5.14-15&1.93-14&6.90-14&5.72-15&3.09-15&6.74-16&1.06-13&2.42-14&8.37-16&1.09-14&5.13-16\\
7.6&5.16-15&1.94-14&6.95-14&5.76-15&3.11-15&6.79-16&1.07-13&2.44-14&8.44-16&1.10-14&5.18-16\\
7.7&4.83-15&1.81-14&6.51-14&5.40-15&2.91-15&6.35-16&1.00-13&2.29-14&7.91-16&1.03-14&4.86-16\\
7.8&4.26-15&1.60-14&5.75-14&4.77-15&2.58-15&5.62-16&8.87-14&2.02-14&7.00-16&9.11-15&4.30-16\\
7.9&3.60-15&1.35-14&4.86-14&4.03-15&2.18-15&4.74-16&7.49-14&1.71-14&5.91-16&7.71-15&3.64-16\\
8.0&2.93-15&1.10-14&3.96-14&3.28-15&1.77-15&3.87-16&6.10-14&1.39-14&4.82-16&6.28-15&2.97-16\\
8.1&2.31-15&8.70-15&3.13-14&2.60-15&1.40-15&3.06-16&4.83-14&1.10-14&3.82-16&4.98-15&2.35-16\\
8.2&1.79-15&6.73-15&2.42-14&2.01-15&1.08-15&2.37-16&3.74-14&8.54-15&2.95-16&3.85-15&1.82-16\\
8.3&1.36-15&5.11-15&1.84-14&1.53-15&8.23-16&1.80-16&2.84-14&6.49-15&2.24-16&2.93-15&1.38-16\\
8.4&1.02-15&3.82-15&1.38-14&1.14-15&6.17-16&1.34-16&2.13-14&4.86-15&1.68-16&2.19-15&1.04-16\\
8.5&7.52-16&2.83-15&1.02-14&8.44-16&4.56-16&9.95-17&1.57-14&3.59-15&1.24-16&1.62-15&7.67-17\\
8.6&5.51-16&2.07-15&7.47-15&6.19-16&3.35-16&7.30-17&1.15-14&2.64-15&9.12-17&1.19-15&5.63-17\\
8.7&4.01-16&1.51-15&5.44-15&4.51-16&2.44-16&5.31-17&8.40-15&1.92-15&6.64-17&8.67-16&4.10-17\\
8.8&2.91-16&1.09-15&3.94-15&3.26-16&1.76-16&3.85-17&6.08-15&1.39-15&4.81-17&6.28-16&2.97-17\\
8.9&2.09-16&7.87-16&2.84-15&2.35-16&1.27-16&2.77-17&4.38-15&1.00-15&3.46-17&4.52-16&2.14-17\\
9.0&1.50-16&5.65-16&2.04-15&1.69-16&9.12-17&1.99-17&3.14-15&7.19-16&2.49-17&3.25-16&1.53-17\\
\br
\end{tabular}
\end{table}

\begin{table}
\caption{Recombination rate coefficients for dielectronic satellite lines
of Ni~XXVII forming Ni~XXIVII (notation: x-y $\rightarrow x\times
10^{-y}$).}
\scriptsize
\begin{tabular}{llllllllllll}
\br
\multicolumn{1}{c}{T(K)}  & \multicolumn{11}{c}{$\alpha_R(cm^3/s)$} \\
 & \multicolumn{1}{c}{o} & \multicolumn{1}{c}{p} & \multicolumn{1}{c}{v} & 
\multicolumn{1}{c}{u} & \multicolumn{1}{c}{r} & \multicolumn{1}{c}{q} & 
\multicolumn{1}{c}{i} & \multicolumn{1}{c}{h} & \multicolumn{1}{c}{t} & 
\multicolumn{1}{c}{s} & \multicolumn{1}{c}{g}\\
\br
6.0&8.63-39&9.75-39&5.48-40&1.20-39&1.90-38&4.00-40&9.66-40&1.49-41&1.71-38&3.84-40&1.13-41\\
6.1&1.90-33&2.15-33&1.25-34&2.79-34&4.80-33&1.05-34&2.58-34&3.97-36&4.64-33&1.06-34&3.12-36\\
6.2&3.10-29&3.50-29&2.08-30&4.75-30&8.77-29&1.97-30&4.90-30&7.55-32&8.96-29&2.06-30&6.08-32\\
6.3&6.40-26&7.24-26&4.40-27&1.01-26&1.98-25&4.54-27&1.14-26&1.76-28&2.12-25&4.91-27&1.45-28\\
6.4&2.57-23&2.90-23&1.79-24&4.17-24&8.52-23&1.99-24&5.04-24&7.77-26&9.43-23&2.20-24&6.49-26\\
6.5&2.79-21&3.15-21&1.97-22&4.63-22&9.80-21&2.32-22&5.92-22&9.12-24&1.11-20&2.62-22&7.73-24\\
6.6&1.08-19&1.22-19&7.70-21&1.82-20&3.96-19&9.47-21&2.43-20&3.74-22&4.60-19&1.09-20&3.20-22\\
6.7&1.83-18&2.07-18&1.32-19&3.13-19&6.96-18&1.68-19&4.33-19&6.67-21&8.24-18&1.95-19&5.75-21\\
6.8&1.61-17&1.83-17&1.17-18&2.79-18&6.32-17&1.54-18&3.97-18&6.12-20&7.59-17&1.80-18&5.31-20\\
6.9&8.48-17&9.59-17&6.18-18&1.48-17&3.40-16&8.30-18&2.15-17&3.31-19&4.12-16&9.80-18&2.89-19\\
7.0&2.95-16&3.34-16&2.16-17&5.17-17&1.20-15&2.95-17&7.66-17&1.18-18&1.47-15&3.51-17&1.03-18\\
7.1&7.40-16&8.36-16&5.43-17&1.30-16&3.06-15&7.53-17&1.96-16&3.02-18&3.77-15&8.99-17&2.65-18\\
7.2&1.43-15&1.62-15&1.05-16&2.53-16&5.98-15&1.48-16&3.85-16&5.92-18&7.42-15&1.77-16&5.22-18\\
7.3&2.25-15&2.54-15&1.66-16&3.99-16&9.49-15&2.35-16&6.12-16&9.43-18&1.18-14&2.82-16&8.32-18\\
7.4&3.00-15&3.39-15&2.22-16&5.34-16&1.27-14&3.16-16&8.25-16&1.27-17&1.59-14&3.81-16&1.12-17\\
7.5&3.52-15&3.97-15&2.60-16&6.27-16&1.50-14&3.73-16&9.73-16&1.50-17&1.88-14&4.50-16&1.33-17\\
7.6&3.71-15&4.20-15&2.75-16&6.63-16&1.59-14&3.96-16&1.03-15&1.59-17&2.00-14&4.79-16&1.41-17\\
7.7&3.61-15&4.08-15&2.67-16&6.46-16&1.56-14&3.87-16&1.01-15&1.56-17&1.96-14&4.68-16&1.38-17\\
7.8&3.29-15&3.72-15&2.44-16&5.89-16&1.42-14&3.54-16&9.24-16&1.42-17&1.79-14&4.29-16&1.26-17\\
7.9&2.85-15&3.22-15&2.11-16&5.10-16&1.23-14&3.07-16&8.02-16&1.24-17&1.55-14&3.72-16&1.10-17\\
8.0&2.36-15&2.67-15&1.75-16&4.24-16&1.02-14&2.55-16&6.68-16&1.03-17&1.29-14&3.10-16&9.14-18\\
8.1&1.90-15&2.15-15&1.41-16&3.41-16&8.24-15&2.05-16&5.37-16&8.28-18&1.04-14&2.49-16&7.36-18\\
8.2&1.49-15&1.68-15&1.10-16&2.67-16&6.46-15&1.61-16&4.21-16&6.49-18&8.16-15&1.96-16&5.77-18\\
8.3&1.14-15&1.29-15&8.46-17&2.05-16&4.96-15&1.24-16&3.23-16&4.98-18&6.27-15&1.50-16&4.43-18\\
8.4&8.59-16&9.72-16&6.38-17&1.54-16&3.74-15&9.33-17&2.44-16&3.76-18&4.73-15&1.13-16&3.34-18\\
8.5&6.40-16&7.23-16&4.75-17&1.15-16&2.79-15&6.95-17&1.82-16&2.80-18&3.53-15&8.45-17&2.49-18\\
8.6&4.71-16&5.33-16&3.50-17&8.47-17&2.05-15&5.12-17&1.34-16&2.07-18&2.60-15&6.23-17&1.84-18\\
8.7&3.44-16&3.89-16&2.56-17&6.19-17&1.50-15&3.75-17&9.80-17&1.51-18&1.90-15&4.56-17&1.34-18\\
8.8&2.50-16&2.83-16&1.86-17&4.50-17&1.09-15&2.72-17&7.12-17&1.10-18&1.38-15&3.31-17&9.76-19\\
8.9&1.81-16&2.04-16&1.34-17&3.25-17&7.88-16&1.97-17&5.15-17&7.93-19&9.98-16&2.39-17&7.05-19\\
9.0&1.30-16&1.47-16&9.65-18&2.34-17&5.67-16&1.41-17&3.70-17&5.70-19&7.18-16&1.72-17&5.07-19\\
\mr
 & \multicolumn{1}{c}{f}& \multicolumn{1}{c}{e}& \multicolumn{1}{c}{k}&
\multicolumn{1}{c}{l}& \multicolumn{1}{c}{d}& \multicolumn{1}{c}{c}&
\multicolumn{1}{c}{j}& \multicolumn{1}{c}{a}& \multicolumn{1}{c}{b}&
\multicolumn{1}{c}{m}& \multicolumn{1}{c}{n} \\
\mr
6.0&7.63-41&1.45-38&1.41-38&1.34-38&2.48-39&2.46-39&4.76-38&9.70-39&3.80-40&2.98-39&8.51-40\\
6.1&2.10-35&4.09-33&4.12-33&3.94-33&7.33-34&7.26-34&1.44-32&3.04-33&1.19-34&9.81-34&2.79-34\\
6.2&4.09-31&8.12-29&8.46-29&8.09-29&1.51-29&1.50-29&3.03-28&6.56-29&2.57-30&2.20-29&6.28-30\\
6.3&9.75-28&1.96-25&2.10-25&2.01-25&3.76-26&3.73-26&7.67-25&1.69-25&6.64-27&5.87-26&1.67-26\\
6.4&4.37-25&8.89-23&9.71-23&9.29-23&1.75-23&1.73-23&3.60-22&8.10-23&3.17-24&2.87-23&8.19-24\\
6.5&5.20-23&1.07-20&1.18-20&1.13-20&2.14-21&2.12-21&4.45-20&1.01-20&3.97-22&3.67-21&1.05-21\\
6.6&2.16-21&4.45-19&5.01-19&4.79-19&9.07-20&8.98-20&1.90-18&4.38-19&1.72-20&1.61-19&4.58-20\\
6.7&3.87-20&8.04-18&9.15-18&8.74-18&1.66-18&1.64-18&3.50-17&8.12-18&3.18-19&3.02-18&8.60-19\\
6.8&3.57-19&7.46-17&8.55-17&8.18-17&1.55-17&1.54-17&3.29-16&7.69-17&3.01-18&2.89-17&8.23-18\\
6.9&1.94-18&4.07-16&4.70-16&4.50-16&8.54-17&8.46-17&1.82-15&4.27-16&1.67-17&1.62-16&4.61-17\\
7.0&6.96-18&1.46-15&1.70-15&1.62-15&3.08-16&3.06-16&6.58-15&1.55-15&6.09-17&5.92-16&1.69-16\\
7.1&1.78-17&3.76-15&4.38-15&4.19-15&7.96-16&7.89-16&1.70-14&4.04-15&1.58-16&1.54-15&4.40-16\\
7.2&3.51-17&7.41-15&8.66-15&8.28-15&1.58-15&1.56-15&3.38-14&8.02-15&3.14-16&3.08-15&8.78-16\\
7.3&5.60-17&1.18-14&1.39-14&1.33-14&2.52-15&2.50-15&5.42-14&1.29-14&5.06-16&4.97-15&1.42-15\\
7.4&7.56-17&1.60-14&1.88-14&1.79-14&3.42-15&3.39-15&7.35-14&1.75-14&6.87-16&6.77-15&1.93-15\\
7.5&8.93-17&1.89-14&2.22-14&2.13-14&4.05-15&4.01-15&8.72-14&2.08-14&8.16-16&8.05-15&2.30-15\\
7.6&9.50-17&2.01-14&2.37-14&2.27-14&4.32-15&4.28-15&9.30-14&2.22-14&8.71-16&8.61-15&2.46-15\\
7.7&9.30-17&1.97-14&2.32-14&2.22-14&4.23-15&4.19-15&9.12-14&2.18-14&8.54-16&8.46-15&2.41-15\\
7.8&8.51-17&1.80-14&2.13-14&2.03-14&3.88-15&3.84-15&8.36-14&2.00-14&7.84-16&7.77-15&2.22-15\\
7.9&7.39-17&1.57-14&1.85-14&1.77-14&3.37-15&3.34-15&7.27-14&1.74-14&6.82-16&6.77-15&1.93-15\\
8.0&6.15-17&1.30-14&1.54-14&1.47-14&2.81-15&2.78-15&6.06-14&1.45-14&5.69-16&5.65-15&1.61-15\\
8.1&4.95-17&1.05-14&1.24-14&1.19-14&2.26-15&2.24-15&4.88-14&1.17-14&4.58-16&4.55-15&1.30-15\\
8.2&3.88-17&8.24-15&9.74-15&9.31-15&1.78-15&1.76-15&3.83-14&9.19-15&3.60-16&3.58-15&1.02-15\\
8.3&2.98-17&6.32-15&7.48-15&7.15-15&1.36-15&1.35-15&2.95-14&7.06-15&2.77-16&2.75-15&7.83-16\\
8.4&2.25-17&4.78-15&5.65-15&5.40-15&1.03-15&1.02-15&2.23-14&5.33-15&2.09-16&2.08-15&5.92-16\\
8.5&1.68-17&3.56-15&4.21-15&4.03-15&7.68-16&7.61-16&1.66-14&3.98-15&1.56-16&1.55-15&4.42-16\\
8.6&1.24-17&2.62-15&3.11-15&2.97-15&5.67-16&5.61-16&1.22-14&2.93-15&1.15-16&1.14-15&3.26-16\\
8.7&9.04-18&1.92-15&2.27-15&2.17-15&4.14-16&4.10-16&8.95-15&2.15-15&8.41-17&8.36-16&2.38-16\\
8.8&6.57-18&1.39-15&1.65-15&1.58-15&3.01-16&2.98-16&6.50-15&1.56-15&6.11-17&6.08-16&1.73-16\\
8.9&4.75-18&1.01-15&1.19-15&1.14-15&2.17-16&2.15-16&4.70-15&1.13-15&4.42-17&4.39-16&1.25-16\\
9.0&3.41-18&7.25-16&8.58-16&8.20-16&1.56-16&1.55-16&3.38-15&8.11-16&3.18-17&3.16-16&9.01-17\\
\br
\end{tabular}
\end{table}

The DES rate coefficients have been tabulated over a wide temperature
range to enable interpolation of $\alpha_s$ at practically any
temperature. However, as described above, $\alpha_s$ can also be
obtained for any temperature of interest using the recombination strength 
$S_{\rm RC}(s)$ or $S(s)$ and the temperature factor, $f(T)$. Although 
the approximation of assuming relatively narrow satellite lines is made 
in (\ref{NP2006a} -- \ref{NP2006c}), use of (\ref{NP2006a} -- 
\ref{NP2006c}) or (\ref{convenient}) is simpler and potentially more 
accurate.

\subsection{Comparison of DES line intensities}

\begin{figure}
\begin{center}
\resizebox{110mm}{!}{\includegraphics{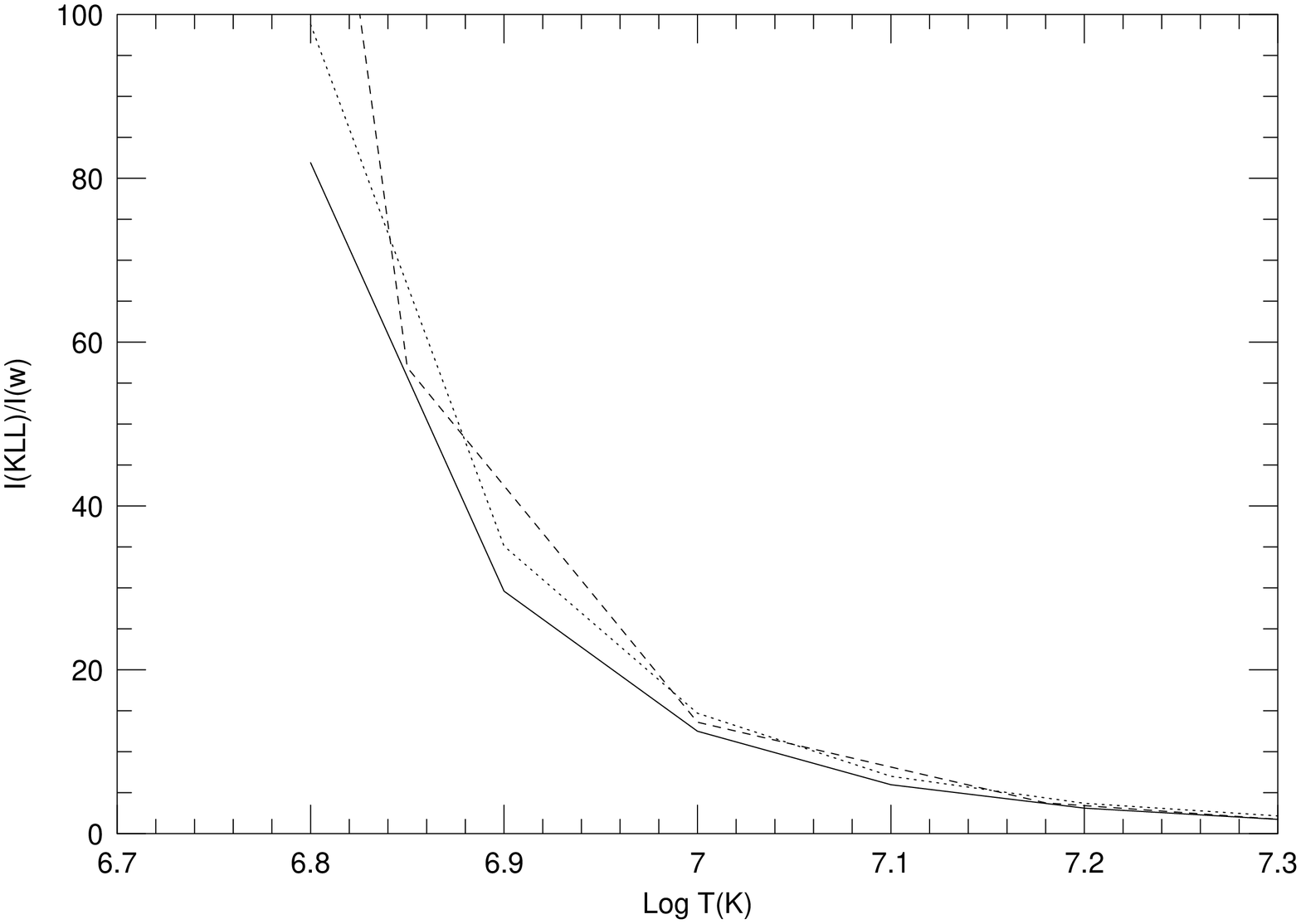}}
\caption{\label{Iratfig}Comparison of intensity ratios $I(KLL)/I(w)$ 
of Fe~XXV with previous calculations: solid line -- present, dashed line
-- Bely-Dubau \etal (1982), dotted line -- Vainshtein and Safronova 
(1978).}
\end{center}
\end{figure}

Given the high temperature sensitivity of individual DES lines, it is
natural to attempt observations that could provide good temperature
diagnostics. The intensity ratios of KLL lines with the $w$-line are 
computed and presented in figure \ref{Iratfig}. We calculated $q_w$
from the electron impact excitation collision strengths for the $w$-line
computed by Pradhan (1985). There is excellent agreement with the $q_w$
values from previous calculations by Bely-Dubau \etal (1982); for
example, $3.22\times 10^{-15}$ and $3.10\times 10^{-15}$ cm$^{-3} s^{-1}$ 
respectively at 10$^7$ K, and $1.23\times 10^{-13}$ and $1.24\times 
10^{-13}$ cm$^{-3} s^{-1}$ at $2\times 10^7$ K. In more recent R-matrix 
calculations, Whiteford \etal (2001) also report very good agreement with 
Pradhan (1985); though their electronically available data lacks the 
finer temperature resolution of Pradhan (1985), Whiteford \etal (2001)'s 
graphical comparisons show negligible differences for dipole allowed 
transitions (within computational uncertainties, usually estimated at 
$<$ 10\% for strong dipole transitions). We have therefore used the 
better resolved (in temperature) calculations of Pradhan (1985) for the 
$w$-line. (We note, however, that there may be significant differences for 
other collisional excitation rates of Fe~XXV).

Except for a few astrophysical observations of the solar corona and 
laboratory experiments such as EBITs, the resolution of individual 
satellite intensities is rare. On the other hand it is easier, and not 
uncommon, to observe the combined intensity of the total KLL DES complex 
(redward of the $w$-line at one extreme). We therefore compute the 
intensity ratio $I(KLL)/I(w)$ as a function of temperature.

We compare present intensity ratios of Fe~XXV (solid curves) in figure 
\ref{Iratfig} with previous results by Vainshtein and Safronova (1978) 
(dashed curve) and Bely-Dubau \etal (1982) (dotted curve) respectively. 
Figure \ref{Iratfig} shows considerable differences in the most sensitive 
low-temperature range, T $< 2\times 10^7$ K ($\log(T) = 7.3$); all
three sets of results appear to converge to good agreement as they approach
the temperature of maximum abundance of Fe~XXV at $\log(T)\sim 7.4$
in coronal equilibrium. But the present values, as well as those of
Bely-Dubau \etal (1982), differ considerably from Vainshtein and
Safronova (1978) all throughout by 20\% to a factor of two or more.
Above $\log(T)\sim 7.2$ our values are in very good agreement with
Bely-Dubau \etal (1982), but show significant differences in the most
sensitive region at about $\log(T)\sim 7.1$. In part, the differences
as well as the agreeement are due to the behavior of the
rate coefficient for the resonance $w$-line which increases with
T, while the DES rate is independent of T. Therefore, the
satellite-to-resonance line ratio at
low temperatures is more reflective of the actual differences between
the various approximations employed to compute the DES resonance
strengths.

\subsection{DES spectra in plasma models}

\begin{figure}
   \begin{center}
     \resizebox{0.8\textwidth}{!}{\includegraphics{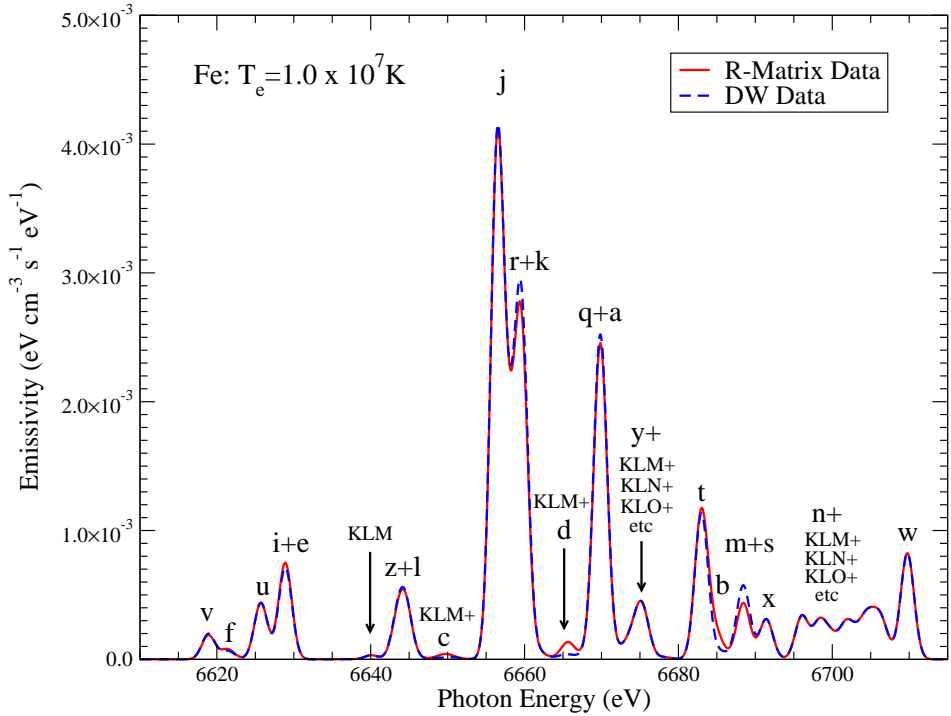}}
      \caption{\label{Fe}Comparison of theoretical dielectronic
satellite spectra of He-like Fe~XXV at T = $10^7$ K using the present 
DES strengths using the unified recombination method (solid, red line) 
and the distorted wave (DW) data from the Los Alamos codes (blue, dashed 
line).}
    \end{center}
\end{figure}

\begin{figure}
   \begin{center}
     \resizebox{0.8\textwidth}{!}{\includegraphics{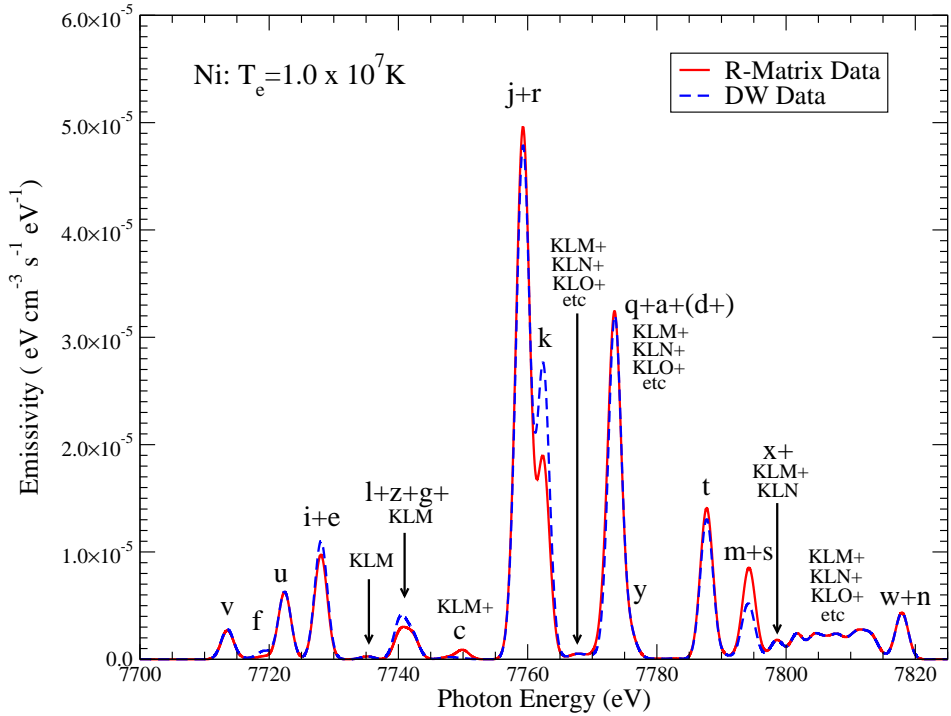}}
      \caption{\label{Ni}Comparison of theoretical dielectronic
satellite spectra of He-like Ni~XXVII at T = $10^7$ K, as in figure \ref{Fe}}
    \end{center}
\end{figure}

Recently, we have modeled the 6.7 keV complex of Fe~XXV and the 7.8 keV
complex of Ni~XXVII in stationary and transient plasma sources using the 
new General Spectral Modeling (GSM) code (Oelgoetz 2006, see also: 
Oelgoetz \etal 2007a,b). GSM is an IRA code that contains the capability 
outlined in (\ref{Aasimple}) such that it can calculate autoionization 
rates ($A_a$) from the unified recombination rates of DES lines. The DES 
spectra for both ions are generated at T =  $10^7$ K and are presented in 
figures \ref{Fe} and \ref{Ni}.

To benchmark the present approach, we compare the calculated Fe~XXV and
Ni~XXVII DES spectra using the present unified satellite rates with 
spectra generated using autoionization rates computed from Los Alamos 
National Laboratory (LANL) suite of atomic physics codes (see: Abdallah 
\etal 1994, 2001). We get very good agreement between the DES spectra 
computed using the two sets of data, as shown in figures \ref{Fe} and 
\ref{Ni}. The calculations for both ions generally follow the approach 
outlined in Oelgoetz \etal (2007a) for the RM data set, as to data 
sources and the complete statistical models. The major exception to this 
is that both models presented use level energies calculated by the LANL 
code CATS (see: Abdallah \etal 1994, 2001) for all energy levels except 
those that give rise to the KLL satellite lines.  The energies for these 
autoionizing states are taken from the data presented in this work. Both 
models use identical energies to facilitate an easy comparison of their 
spectra. The two models differ from each other in that the R-matrix curve
presented here uses autoionization rates calculated from the KLL DES
strengths as computed herein, and the DW data set uses distorted wave 
autoionization rates calculated using the LANL code GIPPER (again, see: 
Abdallah \etal 1994, 2001).

\section{CONCLUSION}

The main conclusions of this work are:

$\bullet$ The theoretical and computational framework for self-consistent
and unified treatment of photoionization and radiative and dielectronic
recombination has now extended the calculation of dielectronic satellite
strengths. Computed DES spectra demonstrate the effect of channel
interference and overlapping profiles, though the practical effect on
rate coefficients is small.

$\bullet$ Theoretical formulation for autoionization rates from the
unified satellite line strengths is given for applications in 
astrophysical models based on IRA methodologies. The first comparison 
for Fe~XXV and Ni~XXVII spectrum shows excellent agreement with data 
from the Los Alamos National Laboratory suite of atomic physics codes.

$\bullet$ Recombination rate coefficients are presented for all 22 KLL
DES lines of Fe~XXV and Ni~XXVII for a wide temperature range. We find
considerable differences with earlier works of more than 20\% at low
temperatures $<$ 10$^7$K, but the agreement improves to $<$ 10\%
towards the temperature of maximum abundance for both ions. While we
expect the present rate coefficients to be more accurate, a conservative
estimate of uncertainties in the temperature range of practical
applications, T$>$ 10$^7$K, is about 10-20\%.

$\bullet$ Intensity ratios $I(s)/I(w)$ of the DES to $w$-line of He-like
Fe and Ni are theoretically computed for practical applications in
high-temperature X-ray plasmas.

$\bullet$ Although the present unified dielectronic resonance strengths
are in good agreement with those computed in previous works using the
IRA, some significant differences are found in temperature ranges where 
the DES are most temperature sensitive, below the temperature of maximum 
abundance of He-like ions in coronal equilibrium.


\subsection{Acknowledgments}
This work was supported partially by the NASA Astrophysical Theory
Program and the Space Astrophysical Research and Analysis programs as 
well partially conducted under the auspices of the United States Department 
of Energy at Los Alamos National Laboratory.  Much of the computational 
work was carried out at the Ohio Supercomputer Center in Columbus, Ohio.

\section*{References} 
 
\def\amp{{\it Adv. At. Molec. Phys.}\ } 
\def\apj{{\it Astrophys. J.}\ } 
\def\apjs{{\it Astrophys. J. Suppl. Ser.}\ } 
\def\apjl{{\it Astrophys. J. (Letters)}\ } 
\def\aj{{\it Astron. J.}\ } 
\def\aa{{\it Astron. Astrophys.}\ }
\def\aas{{\it Astron. Astrophys. Suppl.}\ } 
\def\aasup{{\it Astron. Astrophys. Suppl.}\ } 
\def\adndt{{\it At. Data Nucl. Data Tables}\ } 
\def\cpc{{\it Comput. Phys. Commun.}\ } 
\def\jqsrt{{\it J. Quant. Spectrosc. Radiat. Transfer}\ } 
\def\jpb{{\it Journal Of Physics B}\ } 
\def\pasp{{\it Pub. Astron. Soc. Pacific}\ } 
\def\mn{{\it Mon. Not. R. Astr. Soc.}\ } 
\def\pra{{\it Physical Review A}\ } 
\def\prl{{\it Physical Review Letters}\ } 
\def\zpds{{\it Z. Phys. D Suppl.}\ } 
 
\begin{harvard} 

\item{} Abdallah J, Clark R~E~H, Peek J~M, and Fontes C~J 1994 \jqsrt 
{\bf 51} 1
\item{} Abdallah J, Zhang H~L, Fontes C~J, Kilcrease D~P and Archer 
B~J 2001 \jqsrt {\bf 71} 107
\item Beiersdorfer P, Philips T~W, Wong K~L, Marrs R~E and Vogel 
D~A 1992 \pra {\bf 46} 3812
\item{} Berrington K~A, Burke P~G, Butler K, Seaton M~J, Storey P~J, 
Taylor K~T, and Yan Y 1987 \jpb {\bf 20} 6379
\item{} Berrington K~A, Eissner W and Norrington P~H 1995 \cpc {\bf 92} 290
\item{} Bely-Dubau F, Dubau J, Faucher P and Gabriel A~H 1982 \mn {\bf 198} 239
\item{} Eissner W, Jones M and Nussbaumer H 1974 \cpc {\bf 8} 270
\item{} Gabriel A~H 1972 \mn {\bf 160} 99
\item{} Hummer D~G, Berrington K~A, Eissner W, Pradhan A~K, Saraph H~E 
and Tully J~A 1993 \aa {\bf 279} 298
\item{} Kato T, Fujiwara T and Hanaoka Y 1998 \apj {\bf 492} 822
\item{} Nahar S~N 1996 \pra {\bf 53} 2417
\item{} Nahar S~N 2005 \apjs {\bf 158} 80
\item{} Nahar S~N and Pradhan A~K 1994 \pra {\bf 49} 1816
\item{} Nahar S~N and Pradhan A~K 2004, Review in {\it Radiation Processes 
In Physics and Chemistry} {\bf 70} 323
\item{} Nahar S~N and Pradhan A~K 2006 \pra {\bf 73} 062718-1
\item{} Nahar S~N, Pradhan A~K and Zhang H~L 2000 \apjs {\bf 131} 375
\item{} Nahar S~N, Pradhan A~K and Zhang H~L 2001 \apjs {\bf 133} 255
\item{} Oelgoetz J 2006 Ph.D. thesis, The Ohio State University
\item{} Oelgoetz J, Fontes C~J, Zhang H~L, Montenegro M, Nahar S~N and 
Pradhan A~K 2007a, \mn {\bf 382} 761
\item{} Oelgoetz J, Fontes C~J, Zhang H~L and Pradhan A~K 2007b, \pra 
{\bf 76} 062504-1
\item{} Oelgoetz J and Pradhan A~K 2001, \mn {\bf 327} L42
\item{} Oelgoetz J and Pradhan A~K 2004, \mn {\bf 354} 1093
\item{} Pradhan A~K 1985 \apjs {\bf 59} 183
\item{} Pradhan A~K and Zhang H~L 1997 \jpb {\bf 30} L571
\item{} Sakimoto K, Terao M and Berrington K~A 1990 \pra {\bf 42} 291
\item{} Vainshtein L~A and Safronova U~I 1978 \adndt {\bf 25} 49
\item{} Whiteford A~D, Badnell N~R, Ballance C~P, O'Mullane M~G, Summers 
H~P and Thomas A~L 2001 \jpb {\bf 34} 3179
\item{} Zhang H~L, Nahar S~N and Pradhan A~K 1999 \jpb {\bf 32} 1459
\end{harvard}
\end{document}